\documentclass[12pt]{article}
\usepackage{graphicx,color,amssymb,amsmath}
\begin{document}
\title{\textbf{The heavy quark search at the LHC}} 
\author{B. Holdom%
\thanks{bob.holdom@utoronto.ca}\\
\emph{\small Department of Physics, University of Toronto}\\[-1ex]
\emph{\small Toronto ON Canada M5S1A7}}
\date{}
\maketitle
\begin{abstract}
We explore further the discovery potential for heavy quarks at the LHC, with emphasis on the $t'$ and $b'$ of a sequential fourth family associated with electroweak symmetry breaking.  We consider QCD multijets, $t\overline{t}+\rm{jets}$, $W+\rm{jets}$ and single $t$ backgrounds using event generation based on improved matrix elements and low sensitivity to the modeling of initial state radiation. We exploit a jet mass technique for the identification of hadronically decaying $W$'s and $t$'s, to be used in the reconstruction of the $t'$ or $b'$ mass. This along with other aspects of event selection can reduce backgrounds to very manageable levels. It even allows a search for both $t'$ and $b'$ in the absence of $b$-tagging, of interest for the early running of the LHC. A heavy quark mass of order 600 GeV is motivated by the connection to electroweak symmetry breaking, but our analysis is relevant for any new heavy quarks with weak decay modes.
\end{abstract}

\section{Introduction}
The mystery of the replication of families is part of the flavor problem. But unlike the first three families, a possible fourth family may have a more easily understood role to play. Fourth family fermions with masses above about 550 GeV would couple strongly to the Goldstone bosons of electroweak symmetry breaking \cite{N}. This is another way of saying that these fermions are involved with the strong dynamics of electroweak symmetry breaking. To put it even more strongly, these fermion masses are then the natural order parameters for electroweak symmetry breaking. Meanwhile the fourth family may be the last sequential family and in this way complete the flavor structure of the theory. The joining of these two issues, the flavor problem and electroweak symmetry breaking, is a prime motivation to consider the fourth family.

Strong interactions, rather than a Higgs, would unitarize $WW$ scattering. But given some unknown strong interactions it remains to determine the massive propagating degrees of freedom that most strongly affects this scattering of Goldstone bosons. For example, for the scattering of pseudo-Goldstone bosons of QCD it is the $\rho$. For most theories of electroweak symmetry breaking it is also a boson, either scalar or vector. Instead we are proposing that the propagating degrees of freedom are fermions. This requires that the strong interactions break chiral symmetries without confining the massive fermions. This is reminiscent of the old NJL model, which then forms the basis for a bottom up description of the effective dynamics. In fact in the absence of the Higgs boson, effective four fermion interactions must also be responsible for feeding mass from the heavy fermions to lighter quarks and leptons. The size of such operators are determined by inverse powers of a new mass scale, the scale of flavor physics, which therefore cannot be that far removed from the electroweak scale. Thus a fourth family would not only recast the flavor problem, but it would also force us to conclude that the scale of flavor physics is nearby.

We note that the main effect of a light Higgs boson in electroweak precision data is to shift the value of the $T$ parameter by a positive amount. If there is no light Higgs, then something else must produce a positive $\Delta T$. But the mass splitting in the heavy quark doublet does just that. For a more detailed analysis that shows how the $S$ and $T$ constraints can be satisfied through appropriate masses for the fourth family quark and leptons see \cite{O}. That reference also relates the $t$ mass to a contribution to the heavy quark mass splitting, with the implication that $m_{b'}>m_{t'}$. This result is based on an analysis of the approximate symmetries of operators that may be necessary to account for the $t$ mass while remaining consistent with other constraints such as the $Zb\overline{b}$ coupling. See the appendix for a brief summary of that argument.

Assuming some CKM mixing between the third and fourth families, one or both of the following processes should be important.
\begin{eqnarray}
pp\rightarrow t'\overline{t'}\rightarrow W^+W^-b\overline{b}\label{e1}\\
pp\rightarrow b'\overline{b'}\rightarrow W^+W^-t\overline{t}\label{e2}
\end{eqnarray}
If $m_{b'}$ and $m_{t'}$ differ by more than the $W$ mass then certainly the heavier of $t'$ or $b'$ will decay into the lighter, and only one of these processes will be important. If $m_{b'}$ and $m_{t'}$ differ by less than the $W$ mass then the mass splitting and the value of the CKM mixing angles will determine the importance of transitions between $t'$ and $b'$ involving virtual $W$'s. For example if $m_{t'}=600$ GeV and $m_{b'}=(670, 650, 630)$ GeV then the rate for $b'\rightarrow Wt$ will be comparable to or dominate $b'\rightarrow W^{(*)}t'$ for a mixing angle $\gtrsim$ (.01, .04, .001) respectively. Thus we see how process (\ref{e2}) could still be important even if we are correct about $m_{b'}>m_{t'}$.

In our previous work \cite{F} we developed a search strategy for process (\ref{e1}), where we used the invariant mass of single jets to identify the hadronic decays of $W$'s. This method has been studied in \cite{J} and in cases \cite{U, T} like ours where the $W$'s in the signal events are both well boosted and isolated. The jet invariant mass distribution for the signal events has a strong peak close to $m_W$ for an appropriate cone size in the jet finding algorithm. A $W$-jet is defined to be a jet with invariant mass within $\approx 10$ GeV of $m_W$. It is seen in \cite{F} that this method is significantly less efficient at identifying the less isolated $W$'s of the main irreducible background, $t\overline{t}$ production. Thus in comparison to a more traditional search for heavy quarks \cite{K} where the two jets in $W\rightarrow jj$ are identified, an enhancement of the signal to background ratio $S/B$ in the reconstruction the $t'$ mass is obtained.

In the case of process (\ref{e2}) we shall explore the use of the jet invariant mass technique to identify both $W$'s and $t$'s \cite{J, T} through their hadronic decays. And for both processes (\ref{e1}) and (\ref{e2}) we shall consider a major background not considered in \cite{F}, that of QCD multijets. This background requires a more restrictive event selection. An interesting consequence of these new constraints is that they make possible an effective search without $b$-tagging.

One measure of the background to new heavy particle production is the size of the high energy tail of the $H_T$ distribution (scalar $p_T$ sum of everything in the event including missing energy). The $H_T$ tail is sensitive to initial state radiation, and thus the modeling of ISR in event generators is an important factor in background estimation. For example in stand-alone Pythia\cite{A} and in the case of $t\overline{t}$ production, the default setting has a high cutoff on the phase space of the ISR (``power showers \cite{G}'') in order to obtain realistic $p_T$ distributions of the hardest extra jets. On the other hand realistic $p_T$ distributions of extra jets will also arise from the use of the appropriate perturbative matrix elements, either those that are beyond lowest order at tree level (Alpgen\cite{C}), and/or those that are next-to-leading-order at one-loop (MC@NLO\cite{E}). We found \cite{F} that the generators MC@NLO-Herwig\cite{B}, Alpgen-Herwig and Alpgen-Pythia were in good agreement in their results for the high $H_T$ tail. In comparison stand-alone Pythia with $p_T$-ordered power showers significantly inflates the high $H_T$ tail of $t\overline{t}$ production. (Stand-alone Herwig produced a similar inflation.) The reason for this is that the high cutoff relaxes the relation between different contributions to $H_T$. In particular less energy in the $t\overline{t}$ system, where the partonic cross section is larger, can be made up by the energy of jets from ISR. The improved matrix elements on the other hand more strongly constrain the relative amounts of energy in $t\overline{t}$ versus the extra jets.
\begin{center}\includegraphics[scale=0.32]{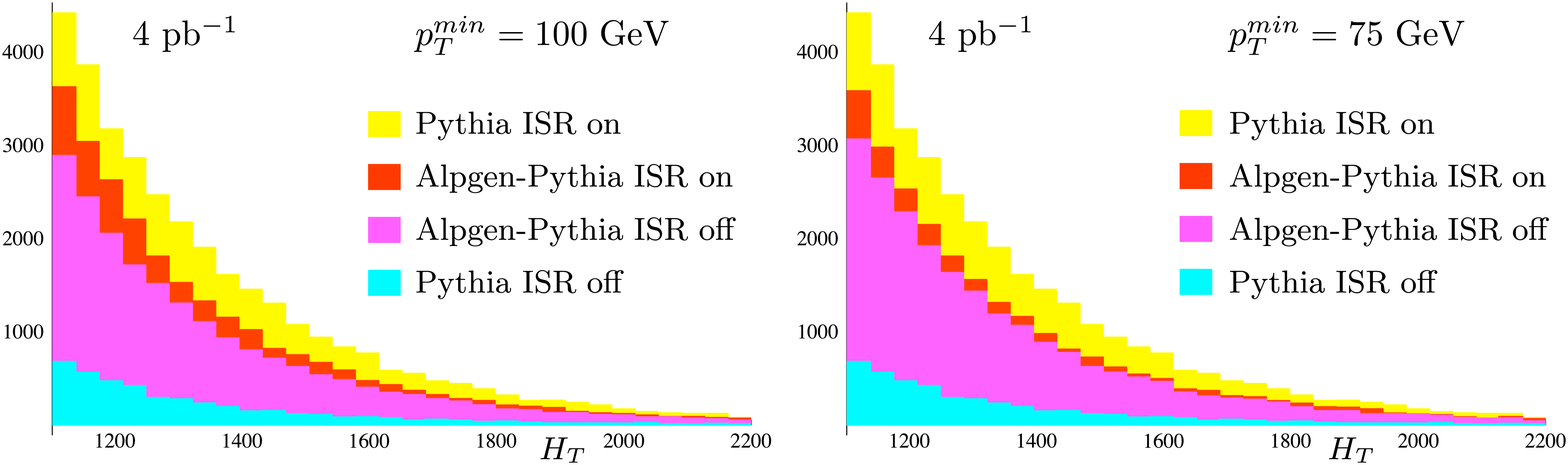}\end{center}
\vspace{-1ex}\noindent Figure 1: Overlayed histograms show the lack of sensitivity of Alpgen-Pythia to Pythia's modeling of initial state radiation in $t\overline{t}$ production, when compared to stand-alone Pythia.  The QW tune is used for both, while $p_T^{min}$ is an Alpgen jet parameter.
\vspace{2ex}

Let us further compare Alpgen-Pythia to stand-alone Pythia, where both are using the same Pythia tune (as described below) with power showers turned off (\verb$MSTP(68)=0$). In Fig.~(1) we show the $H_T$ tails for $t\overline{t}$ production, with and without initial state radiation (\verb$MSTP(61)=1$ or \verb$0$). The very large sensitivity to ISR that is apparent in Pythia is drastically reduced in Alpgen-Pythia. This implies that the MLM jet-parton matching scheme \cite{P} of Alpgen is very efficient at vetoing any extraneous hard ISR from Pythia, beyond that implied by the improved matrix elements of Alpgen. With respect to the Alpgen jet parameter $p_T^{min}$, the slightly greater sensitivity at $p_T^{min}=100$ rather than $p_T^{min}=75$ GeV arises from the $t\overline{t}+0$ jet sample, where the latter becomes more susceptable to Pythia's handling of ISR for larger $p_T^{min}$. Also by comparing the two results of Alpgen-Pythia that include ISR we see very little sensitivity to the choice of $p_T^{min}$, which is a further check of the jet-parton matching scheme.

From this we are encouraged to use Alpgen-Pythia exclusively in the estimation of background. Alpgen currently does not have a user interface for new physics models, and therefore we will use Madgraph\cite{D}-Pythia exclusively for signal generation. We will use the CTEQ6.1 PDF consistently within Alpgen-Pythia and Madgraph-Pythia. This PDF more accurately represents the gluon structure function, which is stronger at the relevent $x$ than given by CTEQ5L. CTEQ6.1 is a NLO PDF, while Alpgen-Pythia only goes part way towards NLO. But if and when NLO matrix elements are introduced it is instructive to see the effect of this while keeping everything else the same, including the PDF. For example the cross section for the production of $t'\overline{t'}$ or $b'\overline{b'}$ (with 600 GeV masses) increases from $\approx0.9$ to $\approx1.4$ pb due to the effect of the NLO matrix elements from MC@NLO. But such enhancements, the K-factors, affect both signal and background and in our previous work we found that Alpgen without K-factors produced a signal to background ratio very similar to MC@NLO.

We will adopt the QW Pythia tune \cite{L} which is basically the popular DW tune adapted to the CTEQ6.1 PDF; only the \verb$PARP(82)$ value is changed. For the renormalization/factorization scale we always choose $\sqrt{\hat{s}}/2$. We note that $S/B$ has little sensitivity to this choice; both signal and background cross sections decrease by nearly identical amounts, about 25\%, when the renormalization scale is increased to $\sqrt{\hat{s}}$.

\section{$t'\overline{t'}$ production and backgrounds}

We use the PGS4 detector simulator \cite{Y} with the ATLAS default set of parameters (from the Madgraph package) and with trigger selections turned off.  We use the cone based jet finder with a cone size of 0.6. We replace the $b$ tag/mistag efficiencies in PGS4 by $(1/2,1/10,1/30)$ (for $|\eta|<2$ and vanishing above) for underlying $b$'s, $c$'s and gluons/light quarks respectively. Our choices should be more appropriate given the high $p_T$'s of the $b$-jets.

For our study of $t'\overline{t'}\rightarrow WWb\overline{b}$ in \cite{F} our focus was on the $t\overline{t}+\rm{jets}$ and $W+\rm{jets}$ backgrounds. The event selection included a lower bound $\Lambda_H$ on the $H_T$ of the event\footnote{In \cite{F} we used the scalar $p_T$ sum of the five hardest objects, which gives similar results.} and a lower bound of $\Lambda_b$ on the $p_T$ of a $b$-tagged jet. We found that $\Lambda_H= 2m_{t'}$ and $\Lambda_b=m_{t'}/3$ worked well, and we will fix $m_{t'}=600$ GeV.\footnote{In \cite{F} we also considered $m_{t'}=800$ GeV.} We require one $W$-jet, defined by having an invariant mass within 9 GeV of $m_W$. In the $t'$ mass reconstruction we consider all pairs of identified $W$ and $b$ jets in each event, where for all such pairs we require an ``angular'' separation $\Delta R<2.5$. We also veto any event with a jet having $|\eta|>2.5$ and $p_T>200$ GeV. 

Our study here will include the QCD multijet background and to adequately suppress this the event selection needs to be tightened further. Thus far we have required one $W$-jet, but now we must require the leptonic decay of the other $W$ and accept the loss of $\approx 80\%$ of the signal. The requirement for isolated leptons and/or missing energy fortunately causes an even more drastic reduction of the multijet background. We consider a loose and a tight cut.
\begin{description}
\item[loose:] isolated lepton \textbf{or} missing energy in excess of 250 GeV
\item[tight:] isolated lepton \textbf{and} missing energy in excess of 30 GeV
\end{description}
The isolated\footnote{The lepton and muon isolation cuts are those described in \cite{W}.} leptons (electron or muon) also have $p_T>20$ GeV.

These new constraints along with the jet mass technique are together so effective that they allow us to treat the $b$-tagging of a jet as optional. Thus our analysis will be done with and without $b$-tags, where in the latter case we maintain the $p_T>200$ GeV constraint on the jet that is combined with the $W$-jet in the $t'$ mass reconstruction. One motivation for eliminating the $b$-tag is to cover the possibility that CKM mixing is such that $t'\overline{t'}\mbox{ (or }b'\overline{b'})\rightarrow W^+W^-q\overline{q}$ is important, where $q$ is a light quark. Another motivation is that $b$-tagging, especially at high $p_T$, may not be very efficient in the early running of the LHC.

For the $t\overline{t}+\rm{jets}$ background we have Alpgen generate samples for 0, 1 and 2 extra hard partons, using the MLM jet-parton matching scheme. The maximum jet pseudorapidity and the minimum jet separation are set to 2.5 and 0.7 respectively. We choose $p_T^{min}=100$ GeV for the Alpgen jet definition; with this choice the $t\overline{t}+1$ jet sample dominates both the exclusive $t\overline{t}+0$ jet sample and the inclusive $t\overline{t}+2$ jet sample in the signal region. More precisely it dominates on the high $H_T$ tails as shown in Fig.~(2a). We are thus ensured that the Alpgen generated matrix elements are controlling the bulk of the showering.
\begin{center}\includegraphics[scale=0.32]{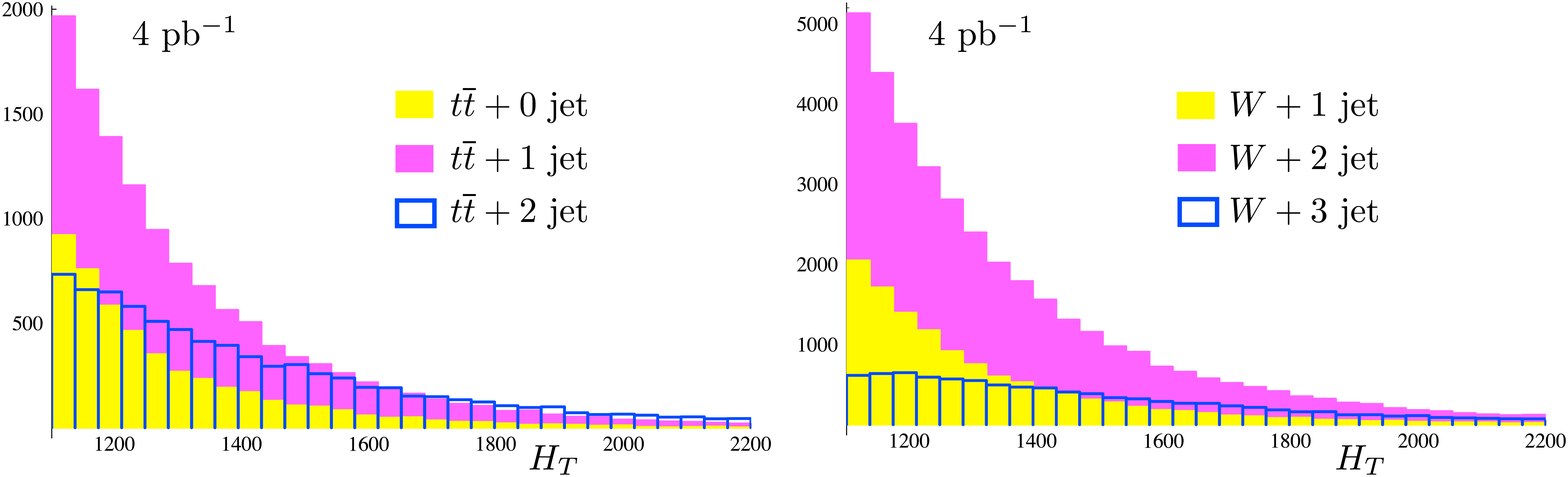}\end{center}
\vspace{-1ex}\noindent Figure 2: Overlayed histograms show the relative sizes of the jet multiplicity samples on the high $H_T$ tail, for a) $t\overline{t}+\rm{jets}$ with $p_T^{min}=100$ GeV (left) and b) $W+\rm{jets}$ with $p_T^{min}=150$ GeV (right).
\vspace{2ex}

For the $W+\rm{jets}$ background we have Alpgen generate the $W+1$, $W+2$, and $W+3$ jet samples, where we allow the $W$ to decay inclusively. Here we use $p_T^{min}=150$ GeV, and we display the relative contributions on the high $H_T$ tail in Fig.~(2b). We also consider the single top production process $pp\rightarrow (t/\overline{t})(\overline{b}/b)W$ since it represents another  irreducible background. This is also modeled in Alpgen with $p_T^{min}=150$ for the $b$-jet. We shall see that this latter background small, and other backgrounds such as $b\overline{b}+\rm{jets}$, $Z+\rm{jets}$, $(W/Z)b\overline{b}$, $(WW/ZZ/WZ)+\rm{jets}$ are even more insignificant.

Potentially more serious is the QCD multijet background. Since the cross sections are so large, it becomes nontrivial for event generators to generate sufficient integrated luminosity to make the background estimate. Here Alpgen again proves helpful since it allows the exclusive 2-jet sample, with its enormous cross section, to be separated out. We will like the 3-jet sample to dominate the 2-jet sample in the signal region, and this occurs if $p_T^{min}$ is not too large. On the other hand by increasing $p_T^{min}$ we can reduce the cross sections, with the reductions proportionally greater for the higher jet multiplicities.  A compromise is to take $p_T^{min}=200$ GeV. This is small enough so that the jets in the 2-jet sample satisfying the $\Lambda_H$ cut are mostly back-to-back so that their combined invariant mass is typically much larger than $m_{t'}$, thus removing them from the signal region. Also this value of $p_T^{min}$ is large enough so that the 4-jet sample, the inclusive sample, is smaller than the 3-jet sample in the signal region. 

The exclusive 2-jet contribution still has an enormous cross section, and so to explore its effect we temporarily drop the lepton/missing energy requirements and the $b$-tagging. We still require a jet with $p_T>\Lambda_b$ to form an invariant mass when combined with a $W$-jet; then we can compare the 2-jet with the sum of the 3-jet and 4-jet samples in the signal region of the $t'$ reconstruction plot. It is easier to generate sufficient events under these conditions and we find that the 2-jet sample is roughly 1/2 as large. Thus we can concentrate on generating sufficient integrated luminosity of the 3 and 4-jet samples with leptons/missing energy/$b$-tagging constraints reinstated, and ignore the 2-jet sample, with the knowledge that the 2-jet sample contributes no more than another 50\%. This possible additional 50\% is certainly an overestimate, since the likelihood of mis-identified leptons or fake missing energy will be less for the 2-jet sample than for the higher multiplicity samples. In fact for the (small) integrated luminosity that we have generated for the 2-jet sample, none of the events survive on the $t'$ mass reconstruction plots. On the plots to follow we do not make any correction for the neglected 2-jet sample.

We show the signal and the various backgrounds as stacked histograms on the $t'$ mass reconstruction plots in Fig.~(3), where the two plots are for the loose and tight lepton/missing energy constraints. We see the successful suppression of the multijet background to almost insignificant levels. Without the lepton/missing energy constraints, the multijet background would be several times higher than the height of the signal peak. We also note that the fall-off of the backgrounds for large invariant mass $M_{Wj}$ is controlled by our constraint on $\Delta R_{Wj}$.
\begin{center}\includegraphics[scale=0.32]{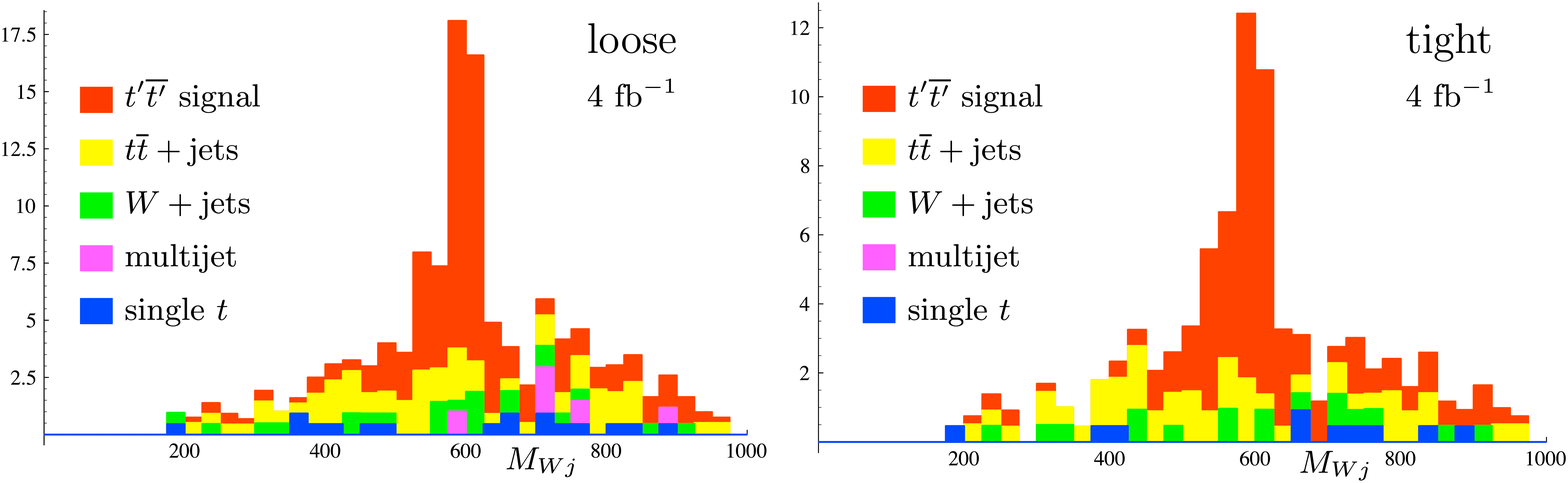}\end{center}
\vspace{-1ex}\noindent Figure 3: The signal from $t'\overline{t'}\rightarrow W^+W^-b\overline{b}$ compared to various backgrounds with $b$-tagging, where the loose and tight cuts refer to the isolated lepton and/or missing energy requirements. The various contributions, including the signal, have been stacked (not overlayed).
\vspace{2ex}

This strength of signal to background encourages us to consider results without the $b$-tag, as shown in Fig.~(4). The multijet background remains small while the $W+\rm{jets}$ background becomes substantially more important. Nevertheless we see that the discovery potential for the heavy quarks is still quite attractive without $b$-tagging, thus providing an opportunity in the early running of the LHC before $b$-tagging methods are well developed.
\begin{center}\includegraphics[scale=0.32]{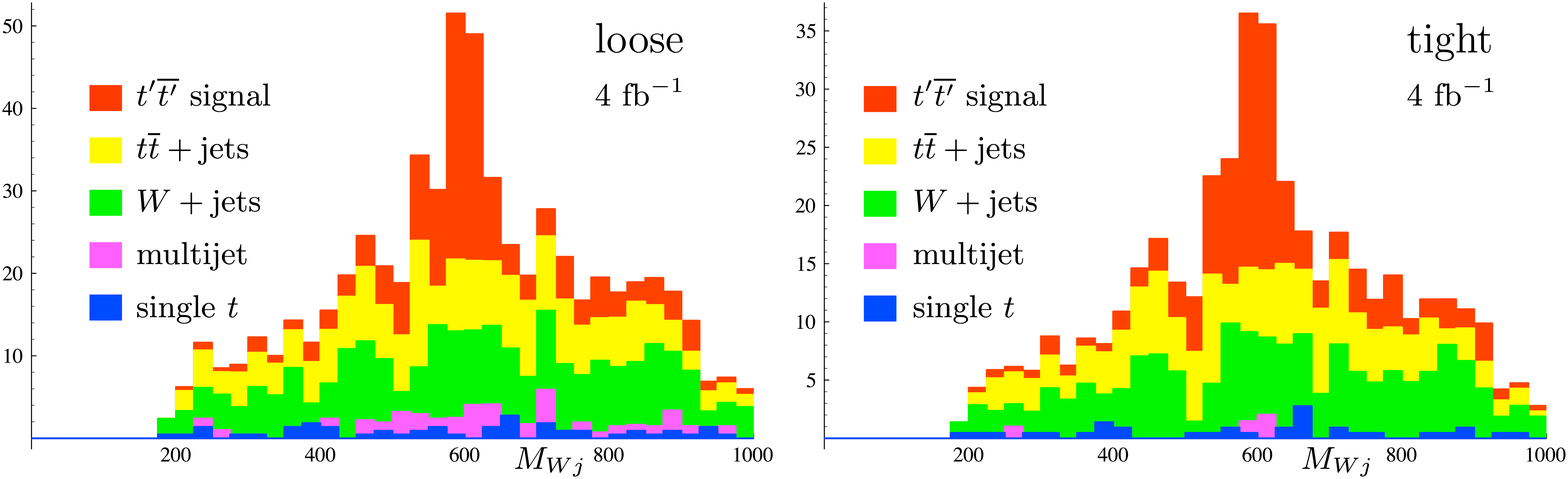}\end{center}
\vspace{-1ex}\noindent Figure 4: The same as the previous figure, without $b$-tagging.
\vspace{2ex}

\section{$b'\overline{b'}$ production and backgrounds}
If $b'$ is larger than the $t'$ mass by more than the $W$ mass, then the following process will occur.
$$pp\rightarrow b'\overline{b'}\rightarrow W^+W^- t'\overline{t'}\rightarrow W^+W^-W^+W^-b\overline{b}$$
This basically increases the signal discussed in the last section. Two of the $W$'s will be relatively soft since the heavy quark mass splitting cannot be too large. The leptonic decays of these $W$'s will add to the likelihood of observing isolated leptons, thus enhancing this contribution to the signal. For $m_{b'}=700$ and $m_{t'}=600$ GeV we find that $b'\overline{b'}$ production has a cross section about 40\% that of $t'\overline{t'}$ production. We compare the two contributions to the signal in Fig.~(5).
\begin{center}\includegraphics[scale=0.32]{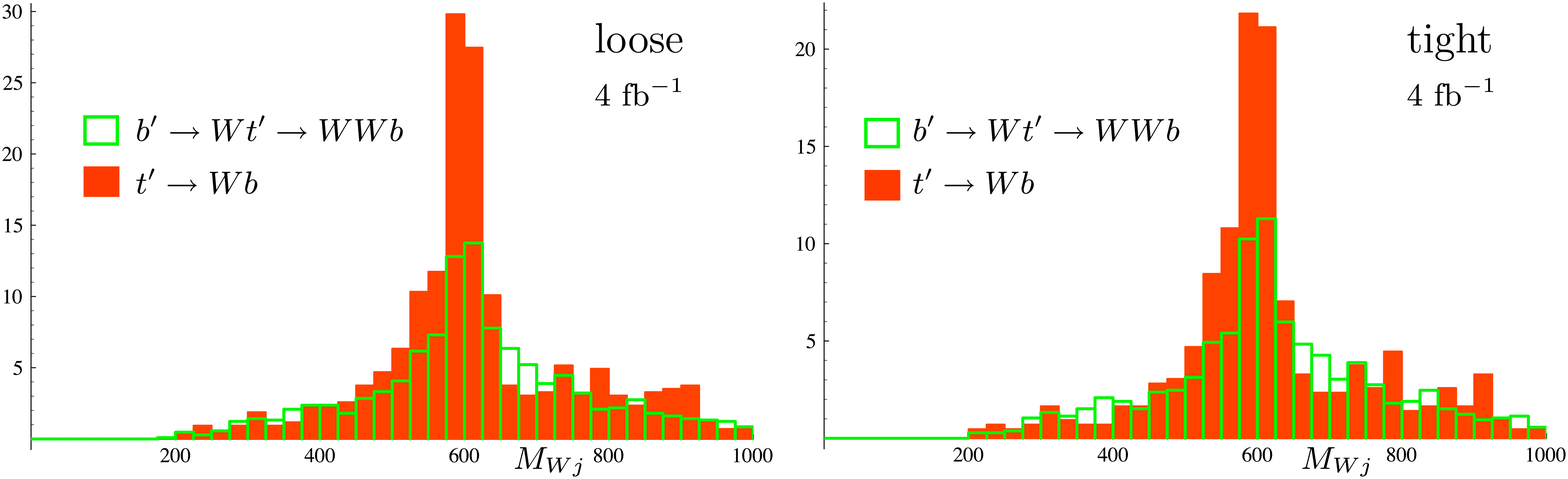}\end{center}
\vspace{-1ex}\noindent Figure 5: The signal appearing in Figure (4) is shown in isolation, along with the additional signal that would arise from $b'\overline{b'}$ production when $m_{b'}=700$ GeV.
\vspace{2ex}

We now consider the process of interest if $b'\rightarrow Wt$ is the dominant decay mode of $b'$:
$$pp\rightarrow b'\overline{b}\rightarrow W^+W^-t\overline{t}$$
Here we set $m_{b'}=600$ GeV. If $m_{t'}$ is sufficiently larger than $m_{b'}$ then this signal is enhanced further through
$$pp\rightarrow t'\overline{t'}\rightarrow W^+W^- b'\overline{b'}\rightarrow W^+W^-W^+W^-t\overline{t},$$
(and the signal of the last section disappears) but we will ignore this in the following. Our object will be to explore the feasibility of using single jet invariant masses to identify both the $t$ and the $W$ through their hadronic decays, and from them reconstruct the $b'$ mass.

A drawback is that the cone size that is optimal to identify $W$ jets is not optimal to identify the $t$ jet, since a significantly larger cone size is necessary to capture the three proto-jets of a boosted $t$ decay.\footnote{The jet mass technique was used in \cite{T} to identify $t$'s from the decay of vector-like quarks more massive than ours, so that the $t$'s were more strongly boosted.} Our compromise, not optimal for either identification, is the choice of 0.8 for the cone size. A $W$-jet is defined by an invariant mass within 9 GeV of $m_W$ as before. For the $t$ the associated invariant mass peak in the signal events is broad and not nearly as strong as the $W$ peak. Thus we make a loose definition of a $t$-jet as a jet with invariant mass greater than 100 GeV and $p_T>300$ GeV.  We use the same loose and tight lepton/missing energy constraints as described before. The only other difference is to tighten the upper bound on $\Delta R$ between the $t$ and $W$ jets to 2.0.

For the QCD jet background we use Alpgen to generate the 2, 3, and 4-jet samples as before. Again none of the 2-jet sample actually generated survives as background to the $b'$-mass reconstruction. We can also bound the possible contribution of a 2-jet sample as before by removing the lepton/missing energy requirements, in which case the 2-jet sample is about 1/4 the size of the $3+4$-jet sample in the signal region. Once again we make no correction for dropping the 2-jet sample.

Using this event selection we produce the results for the $b'$ mass reconstruction in Fig.~(6). Although the signal size is hampered as we have described, we note that the background reduction appears again to be very effective. And it is again of interest to note that our use of the jet mass technique has made possible a search for $b'$ without the use of $b$-tags.
\begin{center}\includegraphics[scale=0.32]{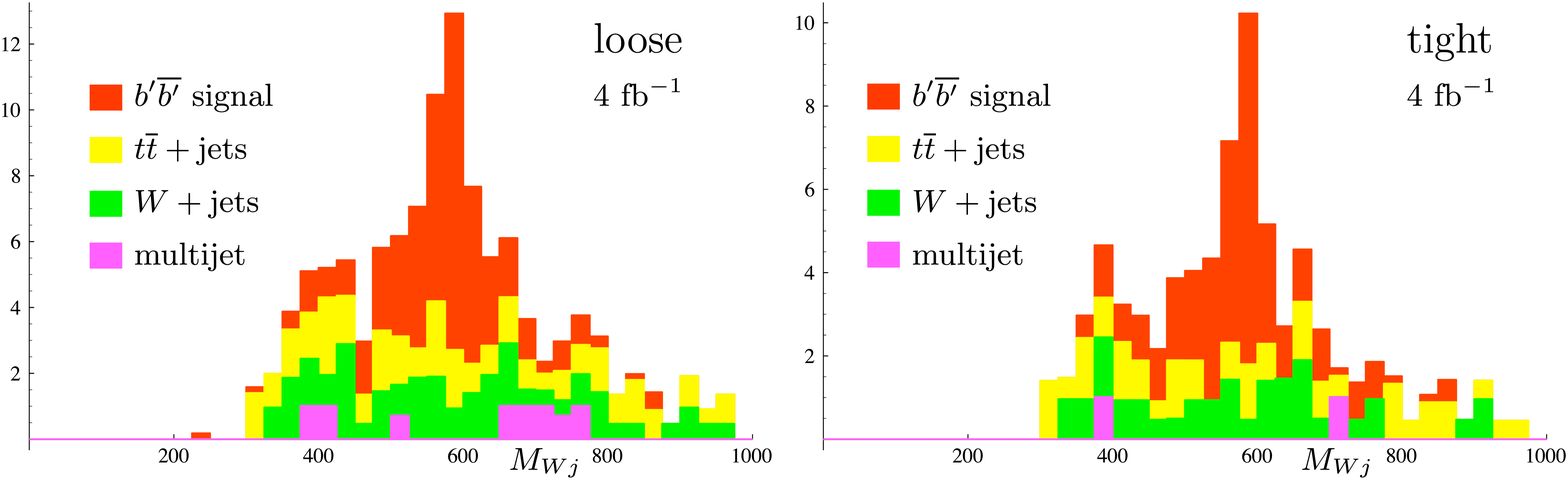}\end{center}
\vspace{-1ex}\noindent Figure 6: The signal from $b'\overline{b'}\rightarrow W^+W^-t\overline{t}$ compared to various backgrounds using the jet mass technique to identify both $W$'s and $t$'s.
\vspace{2ex}

\section{Conclusion}
We believe that our search strategy for new heavy quarks at the LHC improves on the more traditional analysis modeled after the $t$ quark discovery at Fermilab. A key role is played by the jet mass technique to identify $W$'s (and $t$'s) through their hadronic decays, which in the case of $t'\overline{t'}$ production acts to suppress the main irreducible background from $t\overline{t}$ production.  We have found that among the various background processes, only the QCD multijet background forces requirements for isolated leptons and/or missing energy. But with these requirements the search for both the $t'$ and $b'$ can be undertaken without the use of $b$-tagging.

For the actual estimation of backgrounds we have found Alpgen-Pythia to be useful, both to avoid the excessive sensitivity of stand-alone Pythia to the modeling of initial state radiation, and in the estimation of the multijet background. It is possible that our use of the fast detector simulator PGS4 could be leading to an overly optimistic estimate of the background reduction. A full detector simulation is certainly warranted, especially with regard to the efficiency of isolated lepton and missing energy constraints on event selection. Nevertheless the strong signal to background results that we have exhibited provides reason to believe that fourth family quarks could be discovered in the early running of the LHC.

\section{Appendix}
A serious issue for a model of dynamical electroweak symmetry breaking is the generation of the large $t$ mass in a manner compatible with electroweak precision measurements. We briefly summarize an argument \cite{O,R} based on approximate symmetries of effective operators that suggests a way out. Since we are interested in approximate symmetries that can constrain operators that generate mass and/or feed down mass, the approximate symmetries should be axial-like. For the third and fourth family quarks $(q'_L,q'_R,q_L,q_R)$ with $q'=(t',b')$ and $q=(t,b)$, there are two such axial-charge generators to consider: $\cal{Q}$: $(+,-,-,+)$ and $\tilde{\cal{Q}}$: $(+,-,+,-)$.

We then categorize some effective operators of interest in terms of the charges they carry, where all may be written in an $SU(2)_L\times U(1)$ invariant manner. 
\begin{itemize}
\item[1)]  $\overline{t}'_Lt'_R\overline{t}'_Rt'_L\quad\overline{b}'_Lb'_R\overline{b}'_Rb'_L$ (neutral under both charges)
\item[2)] $\overline{t}'_L t'_R\overline{t}_R t_L\quad\overline{b}'_L b'_R\overline{b}_R b_L$ (charged under $\cal{Q}$)
\item[3)] $\overline{b}'_L b'_R\overline{t}_L t_R\quad\overline{t}'_L t'_R\overline{b}_L b_R$ (charged under $\tilde{\cal{Q}}$) 
\end{itemize}
Operators  of type 1 and 2 are such that they can be generated by gauge boson exchange, while the type 3 operators with their $LRLR$ structure cannot be. Type 1 operators represent the dynamics generating mass for the $t'$ and $b'$ while type 2 and 3 operators can feed mass from the fourth family to the third family quarks. Type 2 operators are usually considered for this task. The trouble is that this set of operators includes other operators that are dangerous, in particular those that contribute to the $T$ parameter and the $Zb\overline{b}$ vertex. In particular it is nontrivial to arrange gauge boson exchanges to generate the $t$ mass while not also generating unwanted effects \cite{V,M}.

Type 2 operators are all suppressed if $\cal{Q}$ corresponds to a good approximate symmetry. If $\tilde{\cal{Q}}$ is more badly broken than $\cal{Q}$ then the $t$ mass can instead arise from an operator of type 3. We will refer to $\overline{b}'_L b'_R\overline{t}_L t_R$ as the $t$-mass operator. The $b$-mass on the other hand can come either from the accompanying operator in class 3 (related by a $SU(2)_R$ transformation of the $t$-mass operator) or from an operator of the suppressed class 2. In either case we see that the nonperturbative dynamics responsible for type 3 operators must badly break $SU(2)_R$.
\begin{center}\includegraphics[scale=0.4]{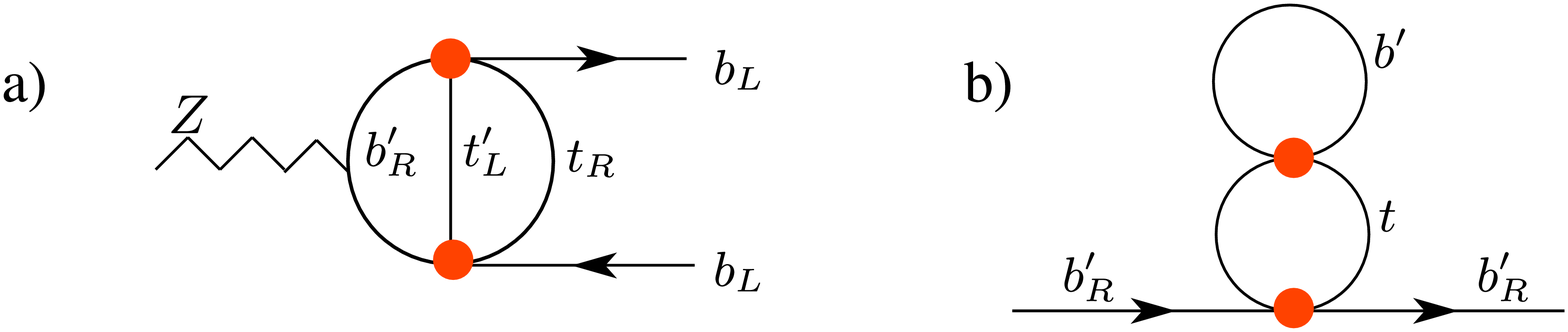}\end{center}
\vspace{-1ex}\noindent Figure 7: Effects arising from two insertions of the $t$-mass operator.
\vspace{2ex}

The mere existence of the $t$-mass operator (its partner $\overline{t}'_L b'_R\overline{b}_L t_R$ by $SU(2)_L$ symmetry is implicit) implies that some operators of class 2 will be generated. But since class 2 operators are $\tilde{\cal{Q}}$ invariant, two insertions of the $t$-mass operator are necessary. The resulting effects are thus suppressed by $(m_t/m_{t'})^2$ and a loop factor. One example is the $Zb\overline{b}$ vertex correction in Fig.~(7a). Another is the correction to the $b'$ mass in Fig.~(7b), which is not shared by the $t'$ mass. This is the origin of the expectation that $m_{b'}>m_{t'}$.

\section*{Acknowledgments}
I thank Brian Beare for his comments, especially those that prompted a removal of the $b$-tag. This work was supported in part by the National Science and Engineering Research Council of Canada.

\end{document}